\documentstyle[12pt]{article}
\begin{document}
%%%%%%%%%%%%%%%%%%%%%%%%%%%%%%%%%%%%%%%%%%%%%%%%%%%%%%%%%%%%%%%%%%%%%
\title{Generalized parity transformations in the regularized
  Chern-Simons theory}
\author{C.~D.~Fosco~\footnote{Electronic
    address: fosco@cab.cnea.gov.ar} and
  A.~L{\'o}pez~\footnote{Electronic address:
    lopezana@cab.cnea.gov.ar}
  \\ \\
  {\normalsize\it Centro At\' omico Bariloche - Instituto Balseiro,}\\
  {\normalsize\it Comisi{\'o}n Nacional de Energ\'{\i}a At{\'o}mica}\\
  {\normalsize\it 8400 Bariloche, Argentina.}}
\date{\today}
\maketitle
%====================================================================
\begin{abstract}
\noindent
We study renormalization effects in the Abelian Chern-Simons (CS)
action. These effects can be non-trivial when the gauge field is
coupled to dynamical matter, since the regularization of the UV
divergences in the model forces the introduction of a parity even
piece in the gauge field action. This changes the classical (odd)
transformation properties of the pure CS action.  This effect, already
discussed for the case of a lattice regularization~\cite{berr}, is
also present when the theory is defined in the continuum and, indeed,
it is a manifestation of a more general `anomalous' effect, since it
happens for every regularization scheme.  We explore the physical
consequences of this anomaly.  We also show that generalized, non
local parity transformations can be defined in such a way that the
regularized theory is odd, and that those transformations tend to the
usual ones when the cutoff is removed.  These generalized
transformations play a role that is tantamount to the deformed
symmetry corresponding to Ginsparg-Wilson fermions~\cite{lusch}
(in an even number of spacetime dimensions).
\end{abstract}
\bigskip
%====================================================================
%%%%%%%%%%%%%%%%%%%%%%%%%%%%%%%%%%%%%%%%%%%%%%%%%%%%%%%%%%%%%%%%%%%%%
%%%%%%%%%%%%%%%%%%%%%%%%%%%%%%%%%%%%%%%%%%%%%%%%%%%%%%%%%%%%%%%%%%%%%
%%%%%%%%%%%%%%%%%%%%%%%%%%% Introduction %%%%%%%%%%%%%%%%%%%%%%%%%%%%
%%%%%%%%%%%%%%%%%%%%%%%%%%%%%%%%%%%%%%%%%%%%%%%%%%%%%%%%%%%%%%%%%%%%%
%%%%%%%%%%%%%%%%%%%%%%%%%%%%%%%%%%%%%%%%%%%%%%%%%%%%%%%%%%%%%%%%%%%%%
\section{Introduction}
It is a well-known fact that the regularization of the UV infinities
of a quantum field theory may break some of the symmetries of the
underlying classical theory. When this breaking is unavoidable, i.e.,
when it cannot be escaped just by choosing a suitable regularization,
one has an `anomaly' in the corresponding symmetry~\cite{anom}.  These
anomalies have very important consequences, ranging from the
constraints for model-building that follow from requiring anomaly
cancellation, to the exact solution of $1+1$ dimensional
models by performing decoupling transformations in a path
integral.

In the present letter, we shall study an anomalous effect that occurs
for a CS theory coupled to a matter field in $2+1$ dimensions.  This
effect amounts to a change in the classical behaviour of the action,
which originally is purely odd under parity, to the `mixed' behaviour
of a sum of two terms with opposite parities. In fact, the general
situation regarding this effect may be thought of as the replacement
of the CS action by a Maxwell Chern Simons (MCS) like action. The use
of a MCS action~\cite{mcs} at intermediate steps, considering the `CS
limit' at the end of the calculation, is of course a well known and
extensively used procedure. We are here, however, considering it as a
regularization, and studying it from the point of view of the
symmetries of the quantum theory.

The replacement of the CS action by the MCS one affects, as we shall
see, the structure of the renormalized effective action in certain
models in such a way that, when the regulator is removed, a non
vanishing anomalous effect remains. The case of the pure CS action
minimally coupled to a Dirac field is dealt with in some detail, since
this is perhaps the most natural system where this phenomenon shows
up. On the other hand, this is a model that has been extensively
studied because of its many interesting properties regarding, for
example, the proper definition of relativistic anyon field operators.

The breaking of the classical parity odd behaviour of the system was
analyzed for the lattice theory in~\cite{berr}, where it was shown
that this breaking will happen for {\em any\/} sensible definition of
the lattice CS action.  It was also suggested in \cite{berr} that the
lattice theory may, perhaps, verify a Ginsparg-Wilson~\cite{gwr} like
relation.  Indeed, the situation is, in more than one aspect, similar
to the breaking of chiral symmetry on the lattice. For the Dirac operator
in odd dimensions, the
corresponding Ginsparg-Wilson relation and their related
generalized  parity transformations have been constructed
on the lattice~\cite{biet} and in the continuum~\cite{cicc}.
Moreover, it has also been shown \cite{biet}  how those
properties can be understood by dimensional reduction from even
dimensions.
The existence of a Ginsparg-Wilson like relation for the CS action
would mean that, although the naive classical transformation
properties of the gauge field are spoiled, there could exist a more
subtle transformation involving the lattice operator, generalizing the
parity odd nature of its continuum version.  In the case of
Ginsparg-Wilson fermions, the generalized chiral transformations are
the ones discovered by L\"{u}scher~\cite{lusch}.  In this article, we
show that a similar phenomenon occurs here for the parity
transformations of the gauge field in the regularized CS action,
and we apply it to the derivation of some consequences.

The organization of this paper is as follows: In
section~\ref{sec:cont}, we consider a Chern-Simons gauge field coupled
to a Dirac field, in the continuum. We show that the regularization
procedure naturally leads one to consider a Maxwell-Chern-Simons (MCS)
theory rather than a pure CS one, hence breaking the classical odd
transformation properties of the Chern-Simons field under parity.
There is, however, a remnant of the classical behaviour which is
manifested by the existence of a generalized parity transformation
under which the MCS action is still odd. This is the content of
section~\ref{symm}, which presents the symmetry transformations
associated with the Ginsparg-Wilson like relation suggested
in~\cite{berr}.  In section~\ref{sec:lat} we show that essentially the
same symmetry holds true for the lattice Chern-Simons theory. In
section~\ref{theor}, we apply the generalized symmetry to the
derivation of some general relations, valid both for the continuum and
lattice versions of the system, and present our conclusions.

%%%%%%%%%%%%%%%%%%%%%%%%%%%%%%%%%%%%%%%%%%%%%%%%%%%%%%%%%%%%%%%%%%%
%%%%%%%%%%%%%%%%%%%%%%%%%%%%%%%%%%%%%%%%%%%%%%%%%%%%%%%%%%%%%%%%%%%
%%%%%%%%%%%%%%%%%%%%%%% Continuum Theory %%%%%%%%%%%%%%%%%%%%%%%%%%
%%%%%%%%%%%%%%%%%%%%%%%%%%%%%%%%%%%%%%%%%%%%%%%%%%%%%%%%%%%%%%%%%%%
%%%%%%%%%%%%%%%%%%%%%%%%%%%%%%%%%%%%%%%%%%%%%%%%%%%%%%%%%%%%%%%%%%%
\section{Continuum theory}\label{sec:cont}
In order to explore the possible renormalization effects for a
dynamical Chern-Simons field, we shall consider in this section a
model consisting of a Dirac field minimally coupled to an Abelian
Chern-Simons gauge field. The generating functional of complete
Green's functions is:
\begin{equation}
  \label{eq:defz}
  {\mathcal Z}[j_\mu, {\bar\eta}, \eta]\;=\;
  \int [{\mathcal D}A_\mu ] {\mathcal D}{\bar\psi} {\mathcal D}\psi \;
  \exp \left\{ i \int d^3x [{\mathcal L}\,+\, j^\mu A_\mu \,+\,
 {\bar\eta} \psi \,+\, {\bar\psi}\eta ]\right\}
\end{equation}
where
\begin{equation}
{\mathcal L}\;=\; {\mathcal L}_F \,+\, {\mathcal L}_{CS}
\end{equation}
\begin{equation}
{\mathcal L}_F \;=\; {\bar \psi} (i \not \! \partial - e \not \!\! A - m)
\psi \;,
\end{equation}
and
\begin{equation}
{\mathcal L}_{CS} \;=\; \frac{\kappa}{2} \epsilon^{\mu\nu\lambda}
A_\mu \partial_\nu A_\lambda
\end{equation}
is the Chern-Simons Lagrangian. In our conventions, the fermionic
fields and $\kappa$ have the dimensions of a mass, while the gauge
field and $e$ of a $({\rm mass})^{1/2}$. $[{\mathcal D}A_\mu]$ denotes
the gauge field functional integration measure, including gauge fixing
factors.

The need for a regularization is then clear, from the evaluation of
the superficial degree of divergence $\omega(G)$ of a proper diagram
$G$ in the perturbative expansion of the generating functional
(\ref{eq:defz}):
\begin{equation}
  \label{eq:wg}
\omega(G) \;=\; 3 - E_F - E_B
\end{equation}
where $E_F$ and $E_B$ denote the number of external fermionic and
bosonic lines in $G$, respectively. This counting corresponds to a
{\em renormalizable\/} theory, and hence it leaves room for the
existence of primitively divergent diagrams. Those are the vacuum
polarization function ($\omega=1$), the fermion self-energy
($\omega=1$), and the vertex function ($\omega=0$). A simple and
convenient gauge invariant regularization scheme for a theory like
this is, of course, the Pauli-Villars method. In this case, it amounts
to replacing ${\mathcal L}$ by a `regularized Lagrangian' ${\mathcal
  L}^{reg}$, defined by:
\begin{equation}
  \label{eq:deflr}
  {\mathcal L}^{reg} \;=\; {\mathcal L}^{reg}_F \,+\, {\mathcal L}^{reg}_{CS}
\end{equation}
where, by following the Pauli-Villars method, the fermion and gauge
field Lagrangians have to be treated differently. For the fermionic
Lagrangian one has to include extra regulator fields that improve the
high momentum behaviour of the fermionic loops. In $2+1$ dimensions,
and for the model we are considering, just one bosonic regulator
${\bar\phi}, \phi$ is sufficient to render all the fermionic loops
convergent:
\begin{equation}
  \label{eq:deflfr}
  {\mathcal L}^{reg}_F\;=\; {\bar\psi} (i \not\! \partial - e \not \!\!A - m ) \psi
+  {\bar\phi} (i \not\! \partial - e \not \!\!A - \Lambda) \phi
\end{equation}
where $\Lambda$ is a mass, proportional to the cutoff of the theory.
Regarding the gauge field, the situation is slightly different, and we
consider it now in more detail. To that end, we shall deal with the
(unregularized) part of ${\mathcal Z}$ which depends on $A_\mu$. It is
clear that this object may be written as follows:
\begin{equation}
  \label{eq:defza}
  {\mathcal Z}_A[J] \;=\; \int {\mathcal D} A_\mu \, \exp \left\{ i \int d^3x
    [{\mathcal L}_{CS}(A) \,+\, J^\mu A_\mu ]\right\} \;,
\end{equation}
where $J^\mu$ denotes the full current to which $A_\mu$ is coupled,
namely,
\begin{equation}
   \label{eq:defJ}
J^\mu \;=\; j^\mu \,-\, e {\bar\psi}\gamma^\mu \psi \,-\,
e {\bar\phi}\gamma^\mu \phi \;.
\end{equation}

The Pauli-Villars method~\cite{itz}, when applied to the gauge field
$A_\mu$, requires the introduction of one {\em massive\/} regulator
field, $B_\mu$, identically coupled to the current, and with a similar
Lagrangian. The mass of $B_\mu$ is also proportional to the cutoff
$\Lambda$.  The regularized version of (\ref{eq:defza}) is then:
\begin{equation}
 \label{eq:defzra}
   {\mathcal Z}^{reg}_A[J] \;=\; \int {\mathcal D} A_\mu \, {\mathcal D}B_\mu
 \;\exp \left\{ i \int d^3x
 [{\mathcal L}^{reg}_{CS}(A,B) \,+\, J^\mu (A_\mu+B_\mu)
 ]\right\} \;,
\end{equation}
where we introduced ${\mathcal L}^{reg}_{CS}$, the `regularized
Chern-Simons Lagrangian', which is defined by:
\begin{equation}
\label{eq:deflrcs}
{\mathcal L}^{reg}_{CS}(A,B) \;=\; {\mathcal L}_{CS}(A) \,-\, {\mathcal L}_{CS}(B)
\,+\,\frac{1}{2} M^2 B_\mu B^\mu \;.
\end{equation}
The idea behind the introduction of the massive field $B_\mu$ is to
improve the large momentum behaviour of the loop integrals which
contain a gauge field propagator without changing the low momentum
behaviour. This is more clearly seen if $B_\mu$ is integrated out in
(\ref{eq:deflrcs}), what can be done easily because the integral is
quadratic. One defines a new field ${\mathcal A}_\mu = A_\mu +
B_\mu$, changes variables from $A_\mu$ and $B_\mu$ to ${\mathcal
  A}_\mu$ and $B_\mu$, and then integrates out $B_\mu$ to obtain:
\begin{equation}
 \label{eq:defzral}
   {\mathcal Z}^{reg}_A[J] \,=\, \int {\mathcal D} {\mathcal A}_\mu \,
 \,\exp \left\{ i \int d^3x
 [-\frac{\xi^2}{4} {\mathcal F}_{\mu\nu}{\mathcal F}^{\mu\nu} \,
 +\, \frac{\kappa}{2} \epsilon^{\mu\nu\lambda}{\mathcal A}_\mu \partial_\nu
 {\mathcal A}_\lambda +J^\mu {\mathcal A}_\mu ]\right\} \,,
\end{equation}
which has a CS form, with $\xi = \frac{\kappa}{M}$ and ${\mathcal
  F}_{\mu\nu}= \partial_\mu {\mathcal A}_\nu - \partial_\nu {\mathcal
  A}_\mu$. This clearly shows that the introduction of the regulator
improves the large momentum behaviour, since the MCS propagator goes
like $k^{-2}$ for large $|k|$, and on the other hand the pure CS
action is recovered when $\xi \to 0$ ($M \to \infty$).

It is interesting to realize that the MCS theory {\em is\/} a
Pauli-Villars regularized version of the CS theory. On the other hand,
it has been known since a long time ago that it is convenient to
evaluate observables in the pure CS theory by starting from the MCS
action, and then to take the $\xi \to 0$ limit at the end of the
calculation.  This approach takes care of many possible sources of
divergences when dealing with the pure CS action coupled to matter.
For example, the celebrated relation between magnetic field
$B=\epsilon_{jk}\partial_j A_k$ and charge $\rho=J^0$
\begin{equation}
\kappa \, B(x) \;=\; - \rho (x)
\end{equation}
of the CS theory is transformed, by the addition of a Maxwell term,
into
\begin{equation}
-\xi^2 \Delta B(x) + \frac{\kappa^2}{\xi^2} B(x) \;=\; -
\frac{\kappa}{\xi^2}\, \rho (x) \;.
\end{equation}
In particular, for a static point-like source, the magnetic flux
becomes
\begin{equation}
B(x)\;\propto\;   K_0 ( M r)
\end{equation}
rather than a $\delta$ function.

The unusual fact here is that the regularized theory is a sensible
physical model, devoided of the unphysical poles usually introduced by
the Pauli-Villars regularization, when one deals with the
regularization of more standard theories.  The reason for this is that
this regularization always adds an extra pole, and the requirement to
improve large momentum behaviour demands the residue at the pole to be
minus the one at the physical singularity. If there is a physical
pole, then the regulator necessarily introduces an unphysical
particle; however, for the pure CS gauge field the physical particle
is missing, and thus the regulator can be chosen to correspond to a
physical pole while improving the UV behaviour of the propagator.

The fact that the purely odd behaviour of $S_{CS}$ is lost is evident,
since the Maxwell action is parity even.

It is worth remarking that, in spite of the fact that the $B_\mu$
field has an explicit mass term, the regularization is gauge
invariant. This is so because the gauge transformations in the
regulated theory are defined by:
\begin{equation}
  \label{eq:gtr}
 A_\mu \;\to\; A_\mu + \partial_\mu \omega \;\;,\;\;  B_\mu \;\to\;
 B_\mu
\end{equation}
which imply ${\mathcal A}_\mu \to {\mathcal A}_\mu + \partial_\mu
\omega$. This shows the consistency of the assumption that $B_\mu$
does not change under gauge transformations, since then the regulated
field ${\mathcal A}_\mu=A_\mu + B_\mu$ transforms in the same way as
$A_\mu$.

\section{Generalized parity transformations}\label{symm}
We shall study here the definition of the parity transformations, both
for the cases of the standard pure CS action, and for the regularized
(MCS) case. The latter covers of course both the regularized CS theory
and a theory defined {\em a priori\/} by a MCS action.

A parity transformation in $2+1$ dimensions is usually defined as a
reflection along only {\em one\/} of the spatial coordinates, since a
spatial inversion ${\vec x} \to -{\vec x}$ is, for a planar system,
equivalent to a rotation (the Jacobian of the coordinate
transformation is equal to $+1$). Thus,
\begin{equation}
  \label{eq:defprt1}
   x_\mu\;\to\;x_\mu^P \;\;:\;\;x^P_0\,=\,x_0 \;,\; x^P_1 \,=\, - x_1
   \;\;,\;\; x^P_2 = x_2
\end{equation}
is a possible definition of a parity transformation of the
coordinates.  The standard, unregularized CS action is odd under these
transformations:
\begin{equation}
  \label{eq:par}
x_\mu \;\to\; x^P_\mu\;\;\;
A_\mu(x) \;\to\; A_\mu^P(x^P) \;=\; \frac{\partial x^P_\mu}{\partial x_\nu}
A_\nu(x) \;,
\end{equation}
since
$$
S_{CS}[A] \;=\; \int d^3x \,\frac{\kappa}{2} \,
\epsilon^{\mu\nu\lambda} A_\mu(x) \frac{\partial}{\partial x_\nu}
A_\lambda (x)
$$
\begin{equation}
  \label{eq:par1}
 \;=\; - \int d^3x^P \,\frac{\kappa}{2} \,
  \epsilon^{\mu\nu\lambda}
A^P_\mu(x^P) \frac{\partial}{\partial x^P_\nu} A^P_\lambda(x^P) \;=\;
- S_{CS}[A^P] \;.
\end{equation}
This action is also odd under time inversion. Thus we can also use a
more symmetric expression for the transformations, introduced
in~\cite{berr}, where `parity' is defined by the full spacetime
inversion:
\begin{equation}
  \label{eq:defprt}
  x \;\to\; x^I \;\;\;\;\;\;\;x^I_\mu\,=\,-x_\mu \;,
\end{equation}
which may be thought of as a time inversion composed with a spatial
rotation in $\pi$. The $I$ transformation has the notational
convenience that it has an identical effect on all the coordinates.
Everything we shall do for the $I$ transformations has of course an
immediate analogy for the $P$ transformations.

In order to study the structure of these transformations and their
generalizations, it is convenient to use a more abstract notation. For
example, the vector field $A_\mu(x)$ and its Fourier transform ${\tilde
  A}_\mu(p)$ will be regarded as the coordinate and momentum
representations, respectively, of some abstract vector field $|A_\mu \rangle$
in a Hilbert space:
\begin{equation}
  \label{eq:not1}
A_\mu (x) \;=\; \langle x | A_\mu \rangle \;\;,\;\;\;
{\tilde A}_\mu (p) \;=\; \langle p | A_\mu \rangle \;.
\end{equation}
Then, for the $I$ transformation acting on $A_\mu$, we have:
$$
\langle x^I | A^I_\mu \rangle \;=\; - \langle x | A_\mu \rangle
$$
\begin{equation}
\label{eq:not2}
\langle x | A^I_\mu \rangle \;=\; - \langle x^I | A_\mu \rangle
\;=\; - \langle x |{\mathcal I}|A_\mu \rangle
\end{equation}
where $\mathcal I$ denotes the operator that performs the $I$ transformation on
the coordinates. Its matrix elements are then
\begin{equation}
   \label{eq:not3}
\langle x |{\mathcal I}| y \rangle \;=\; \langle x^I | y \rangle \;=\;
 \langle x | y^I \rangle \;=\;
\delta^{(3)}(x + y) \;.
\end{equation}
Then the inversion operator ${\mathbf I}$, when acting on $A_\mu$ is
given by:
\begin{equation}
   \label{eq:not4}
{\mathbf I} |A_\mu \rangle \;=\; |A_\mu^I\rangle \;=\; - \,
{\mathcal I} \, |A_\mu \rangle \;,
\end{equation}
or
\begin{equation}
   \label{eq:not5}
{\mathbf I} \;=\; - {\mathcal I} \;.
\end{equation}
With these conventions, and working now with Euclidean spacetime
conventions, the CS action is written as:
\begin{equation}
   \label{eq:not6}
S_{CS}[A]\;=\; \frac{\kappa}{2}\, \langle A_\mu | R_{\mu\nu}
| A_\nu \rangle
\end{equation}
with $R_{\mu\nu}\equiv i \epsilon_{\mu\lambda\nu}\partial_\lambda$,
and its quality of being odd under ${\mathbf I}$ follows from:
\begin{equation}
\label{eq:not7}
S_{CS}[A^I]\;=\; \frac{\kappa}{2}\, \langle A^I_\mu |R_{\mu\nu}
| A^I_\nu \rangle \,=\,
\frac{\kappa}{2}\, \langle A_\mu |\, {\mathcal I} \, R_{\mu\nu} \,
{\mathcal I}\,
|A_\nu \rangle \;=\; - S_{CS}[A]\;,
\end{equation}
which is tantamount to
\begin{equation}
{\mathbf I} R {\mathbf I} \;=\; - R \;\;\Leftrightarrow\;\;
\{{\mathbf I} , R \}\,=\, 0\,,
\end{equation}
where the anticommutativity is derived by using also the obviously
satisfied relation ${\mathbf I}^2 = 1$.

Let us now turn to the MCS action which, in Euclidean spacetime and
with the above conventions can be written as
\begin{eqnarray}
S_{MCS}[A] &=& \frac{1}{2} \langle A_\mu | \left[ \kappa
  R_{\mu\nu}
  \,+\,\frac{\kappa^2}{M^2}(-\partial^2)\delta^\perp_{\mu\nu}\right]
|A_\nu\rangle \nonumber\\
   \label{gp1}
&\equiv &\frac{\kappa}{2}\, \langle A_\mu | {\tilde R}_{\mu\nu} | A_\nu \rangle
\end{eqnarray}
where $\delta^\perp_{\mu\nu}=\delta_{\mu\nu}-
\displaystyle{\frac{\partial_\mu \partial_\nu}{\partial^2}}$ and
\begin{equation}
   \label{deftlr}
{\tilde R} \;=\; R ( \, 1 - \frac{\kappa}{M^2} \, R\, ) \;.
\end{equation}
The generalized inversion transformations ${\mathbf{\tilde I}}$ may be
found by defining a general linear transformation for $|A_\mu\rangle$,
\begin{equation}
   \label{eq:git}
|A_\mu\rangle \to |A^I_\mu\rangle \,=\, {\mathbf{\tilde I}} |A_\mu\rangle
\,=\,- {\mathcal I}\,(f \delta^\perp_{\mu\nu}\,-\,g R_{\mu\nu})
|A_\nu\rangle
\end{equation}
(with ${\mathcal I}$ as defined in (\ref{eq:not3})), and imposing the
condition:
\begin{equation}
   \label{eq:cond}
{\mathbf{\tilde I}}\, {\tilde R}_{\mu\nu} \, {\mathbf{\tilde I}} \;=\; -
{\tilde R}_{\mu\nu} \;.
\end{equation}
\newpage The scalar functions $f$ and $g$ are easily shown to be
\begin{eqnarray}
f &=& \left( 1 - \xi^2 \, \frac{\partial^2}{M^2}\right)^{-\frac{1}{2}}
\nonumber\\
g &=& \frac{\kappa}{M^2} \,
\left(1 - \xi^2 \, \frac{\partial^2}{M^2}\right)^{-\frac{1}{2}} \,.
   \label{eq:fg}
\end{eqnarray}
The transformation law for the longitudinal part of the gauge field is
arbitrary, since it is not determined by the equation (\ref{eq:cond}).
We choose, for simplicity,
\begin{equation}
   \label{eq:lpt}
{\mathbf{\tilde I}} |A_\mu\rangle
\,=\,- {\mathcal I}\,(f \delta_{\mu\nu}\,-\, g R_{\mu\nu})
|A_\nu\rangle \;.
\end{equation}
It is straightforward to check that the transformations defined by
these coefficient functions indeed verify (\ref{eq:cond}), by direct
substitution. However, it is perhaps more instructive to realize that
the generalized inversions defined by (\ref{eq:git}) can also be
written as:
\begin{equation}
   \label{eq:git1}
{\mathbf{\tilde I}} \;=\;-\,{\mathcal I}\, \sqrt{\frac{1 - \xi \,
\frac{1}{M} R}{ 1 + \xi \,
\frac{1}{M} R}}
\end{equation}
where the property (\ref{eq:cond}) is more explicit. We can also use
(\ref{eq:git1}) to show that $\{ {\mathbf{\tilde I}} , {\tilde
  R}_{\mu\nu} \} = 0$, since
$$
{\mathbf {\tilde I}}^2 \,=\,{\mathcal I}\, \sqrt{\frac{1 - \xi \,
    \frac{1}{M} R}{1 + \xi \, \frac{1}{M} R}} \,{\mathcal I}\,
\sqrt{\frac{1 - \xi \, \frac{1}{M} R}{ 1 + \xi \, \frac{1}{M} R}}
$$
\begin{equation}
  \label{eq:idemp}
=\, \sqrt{\frac{1 + \xi \,
\frac{1}{M} R }{1 - \xi \,
\frac{1}{M} R}} \, \sqrt{\frac{1 - \xi \,
\frac{1}{M} R}{1 + \xi \,
\frac{1}{M} R}} \,=\, 1 \;.
\end{equation}

We conclude this section by pointing out that the introduction of
the Maxwell term to regulate the theory has a non-trivial effect on
the renormalized theory. Indeed, the two-particle scattering amplitude
including one-loop effects has a local repulsive interaction which survives
even when taking the `anyon limit' $\xi \to 0$~\cite{kogan}.
%%%%%%%%%%%%%%%%%%%%%%%%%%%%%%%%%%%%%%%%%%%%%%%%%%%%%%%%%%%%%%%%%%%
%%%%%%%%%%%%%%%%%%%%%%%% Lattice Theory %%%%%%%%%%%%%%%%%%%%%%%%%%%
%%%%%%%%%%%%%%%%%%%%%%%%%%%%%%%%%%%%%%%%%%%%%%%%%%%%%%%%%%%%%%%%%%%
\section{Lattice theory}
\label{sec:lat}

We briefly discuss here the meaning of the symmetry presented in
section~\ref{symm} from the point of view of the lattice CS theory.
To that end, we review the result presented in \cite{berr}. In this
work, the authors show that given the CS action:
\begin{equation}
   \label{lcs}
S_{CS}\;=\; \sum_{x,y} A_\mu (x) G_{\mu\nu}(x-y) A_\nu (y)
\end{equation}
and assuming that it is local on the lattice, gauge invariant, and
odd under parity, then (\ref{lcs}) is not integrable. However,
relaxing the last condition, the more general form of a local gauge
invariant action in three dimensions includes a lattice Maxwell term
\begin{equation}
   \label{mx}
S_{M}\;=\; \sum_{x,y} A_\mu (x) M_{\mu\nu}(x-y) A_\nu (y)
\end{equation}
where
\begin{equation}
   \label{maxwell}
 M_{\mu\nu}(x-y)=- \Box  \, \delta_{\mu\nu} + d_\mu {\hat d}_\nu \;.
\end{equation}
In this equation $ \Box = \sum_{\mu =0}^2 d_\mu {\hat d}_\mu$ is the
Laplacian in three dimensions. The forward and backward difference
operators are given by $d_\mu f(x)= f(x + {\hat \mu}) - f(x)=(s_\mu -
1) f(x)$ and ${\hat d}_\mu f(x) = f(x) - f(x - {\hat \mu})(1-
s_\mu^{-1}) f(x)$ respectively, where $s_\mu$ is the forward
translation operator.

It was shown in reference~\cite{berr} that the only gauge invariant
way to regularize the CS action is to add a parity even term such as
the Maxwell term (\ref{maxwell}).  The regularization of the extra
zeroes in the CS action is due to the fact that the Maxwell term opens
up a gap for them. Thus its introduction avoids one of the undesired
features of the lattice CS action, at the price of destroying the odd
behaviour of the pure CS action.  However, we can show that, as in the
continuum case, there is a generalized symmetry that is in fact
preserved.  That the same symmetry of Section~\ref{symm} holds on the
lattice, can be seen for example, from the Fourier version of
(\ref{lcs}) and (\ref{mx}), which is formally identical to its
continuum counterpart, except from the different momentum ranges.

The lattice Fourier transformation of the gauge field $A_{\mu}$ is given
by
\begin{equation}
   \label{ftrans}
 A_\mu (x) = \int_{\cal B} {\frac {d^3p}{(2\pi)^3}} e^{-i px} e^{-i {\frac{p_{\mu}}{2}}}
{\tilde  A}_{\mu}(p) \;,
\end{equation}
where the lattice the integration over momenta is restricted to the
Brillouin zone ${\cal B}$.  Therefore the Fourier
transformation of the Chern-Simons action is
\begin{equation}
   \label{lcsft}
S_{CS}\;=\; \int_{\cal B} {\frac {d^3p}{(2\pi)^3}} {\tilde  A}_{\mu}(p)
{\tilde  G_{\mu\nu}}(p) {\tilde  A}_{\nu}(-p)
\end{equation}
with ${\tilde G_{\mu\nu}}(p)= e^{-i {\frac{p_{\mu}}{2}}} G_{\mu\nu}(p)
e^{i{\frac{p_{\nu}}{2}}}$. It was shown in~\cite{berr} that by requiring
locality on the lattice, parity oddness and gauge invariance, the
kernel ${\tilde G_{\mu\nu}}(p)$ must be of the form:
\begin{equation}
\label{gmn}
{\tilde  G_{\mu\nu}}(p)= i  \epsilon_{\mu\rho\nu} {\hat p}_{\rho} h(p)
\end{equation}
with ${\hat p}_{\rho}=-2i \sin{\frac {p_{\rho}} 2}$, and $h(p)$ an
even analytic function of $p$.

As for the Maxwell term, its Fourier transformation is
\begin{equation}
   \label{mxft}
S_{M}\;=\;{\frac {1}{e^2}}  \int_{\cal B} {\frac {d^3p}{(2\pi)^3}} {\tilde  A}_{\mu}(p)
{\tilde   M_{\mu\nu}}(p) {\tilde  A}_{\nu}(-p)
\end{equation}
being ${\tilde M_{\mu\nu}}(p)= e^{-i{\frac{p_{\mu}}{ 2}}} M_{\mu\nu}(p) e^{i
  {\frac {p_{\nu}}{ 2}}} = -{\hat p}^2 \delta_{\mu\nu}+ {\hat p}_{\mu} {\hat
  p}_{\nu}$.

Using equations (\ref{lcsft}) and (\ref{mxft}) it is simple to check
that the Maxwell Chern-Simons action can be written as
\begin{equation}
   \label{mcsft}
S\;=\; \int_{\cal B} {\frac {d^3p}{(2\pi)^3}} {\tilde  A}_{\mu}(p)
\Gamma_{\mu\nu}(p) {\tilde  A}_{\nu}(-p)
\end{equation}
where
\begin{eqnarray}
\Gamma_{\mu\nu}(p) &=& f(p) \delta_{\mu\nu}^{\perp} (p) + i g(p) Q_{\mu\nu}(p)   \\
f(p) &=& {\frac {4}{e^2}} {\sum_{\alpha=0}^{2} \sin^2{\frac {p_\alpha}{2}}} \\
g(p) &=& 2 h(p)  \sqrt{{\sum_{\alpha=0}^{2} \sin^2{\frac {p_\alpha}{2}}}}  \\
\delta_{\mu\nu}^{\perp} &=& \delta_{\mu\nu} - {\frac {\sin({\frac {p_\mu}{2}})
\sin({\frac {p_\nu}{2}})}{{\sum_{\alpha=0}^{2} \sin^2({\frac {p_\alpha}{2}})}}} \\
Q_{\mu\nu} &=& {\frac {1}{\sqrt{ - {\hat p}^2}}} \epsilon_{\mu\alpha\nu} {\hat p}_{\alpha}
\end{eqnarray}
We can now find the generalized parity transformations following the
same steps as in Section~\ref{symm}. The result for the gauge field is
\begin{equation}
\label{gpt}
A_{\mu}^I(p^I)= i {\frac {f(p)}{\sqrt {f^2(p)+g^2(p)}}} Q_{\mu\rho} A_{\rho}(p) -
{\frac {g(p)}{\sqrt {f^2(p)+g^2(p)}}} \delta_{\mu\rho}^{\perp} A_{\rho}(p)
\end{equation}
Notice that when $f(p) \to 0$, i.e., when the action becomes the
Chern-Simons action, we obtain the usual parity transformation
$A_{\mu}^I(x^I)= -A_\mu (x)$. Therefore , this
is the new parity transformation under which the Chern-Simons action
is still odd but the kernel is integrable, and  whose existence was
suggested in reference~\cite{berr}.

Thus, we conclude that we can use the generalized inversion or parity
transformation (\ref{gpt}) as a substitute for the usual versions of
those discrete symmetries. Indeed, as we shall see in the next
section, the generalized symmetries are more useful than the usual
ones when one tries to derive non-perturbative relations, like index
theorems.

\section{Conclusions}\label{theor}
Let us derive some consequences from the generalized inversion (or parity)
symmetry. We first note that one may use the (algebraic) relations
involving the ${\mathbf I}$ operator and the kinetic operator
${\tilde R}_{\mu\nu}$ to construct an index theorem. Namely, we consider the
quantity $\omega$, the trace (in functional space and Lorentz indices) of
${\mathbf I}$:
\begin{equation}
   \label{it1}
\omega({\mathbf I}) \;=\; {\rm Tr} ( {\mathbf I} )\;.
\end{equation}
Defining the (normalized) eigenvectors of $R$:
\begin{equation}
   \label{eq:eigen}
R_{\mu\nu} \phi^{(n)}_\nu (x) \;=\; \lambda_n \, \phi^{(n)}_\mu
\end{equation}
the anticommutativity of $R$ and ${\mathbf I}$ implies that
the generalized inversion pairs eigenvectors of opposite $\lambda_n$.
Then all except the $\lambda_n=0$ states cancel out in the trace, when it
is evaluated on the basis of the $\phi^{(n)}$:
\begin{equation}
   \label{eq:it2}
\omega({\mathbf I}) \;=\; \sum_{\lambda_n =0}
\langle n | {\mathbf I} | n \rangle \;=\; n_+ - n_- \;,
\end{equation}
where $n_\pm$ denote the number of zero modes of positive and negative
parity, respectively. On the other hand, these zero modes verify
\begin{equation}
  \label{eq:zm1}
\epsilon_{\mu\nu\lambda} \partial_\nu \phi^{(0)}_\lambda (x) \;=\; 0
\end{equation}
and this means of course that $\phi^{(0)}_\lambda$ is locally a `pure gauge'
vector field $\phi^{(0)}_\lambda= \partial_\lambda \omega$. The number of
independent (normalizable) zero modes will thus be identical to the number of
independent holonomies in the spacetime manifold where the action is defined.
This, of course, has been derived in the context of the formal, unregularized
CS action. Assuming now that one evaluates the trace of the standard inversion
operator in the lattice theory, we immediately run into trouble, since there
are spurious zero modes of the CS action that cancel out any non-vanishing
contribution to the index theorem. The situation is of course analogous to the
case of  using the naive Dirac action for massless fermions on the lattice.
However, having the generalized inversion symmetry, we may consider the trace
of ${\mathbf{\tilde I}}$, which anticommutes with the lattice MCS operator
${\tilde R}$, and the non-zero modes are still paired.
The problem with the spurious zero modes is avoided, because now there is a gap
for these modes at the edges of the Brillouin zone.

Let us now show that the introduction of a parity even, Maxwell-like term
in the action is not a special feature of the Pauli-Villars method,
but rather a common feature of any sensible regularization implementable
at the Lagrangian level. With full generality, a general regularized
action $S^{reg}$ can be written as
\begin{equation}
   \label{eq:greg}
S^{reg}[{\mathcal A}] \;=\; S_{PV}[{\mathcal A}] \,+\, S_{PC}[{\mathcal A}]
\end{equation}
where $S_{PV}$ and $S_{PC}$ denote parity violating and conserving terms,
respectively. Obviously, the MCS action is a particular case of this
general form, and it does regulate the gauge field propagator. Of course,
one may think of more general actions corresponding to higher derivative
versions of the MCS theory. For example,
\begin{equation}
   \label{eq:greg1}
S^{reg}[{\mathcal A}]\;=\;
\frac{\kappa}{2} \langle A_\mu | u(\frac{\partial^2}{M^2}) R_{\mu\nu}
| A_\nu \rangle \,+\,\frac{1}{4} \langle F_{\mu\nu}| v(\frac{\partial^2}{M^2})
|F_{\mu\nu}\rangle \;,
\end{equation}
where $u$ and $v$ are functions chosen in order to obtain the desired
behaviour for the resulting propagator. A crucial observation is that,
to preserve gauge invariance, one cannot set $v\,\equiv\,0$. The reason is
that, for $v=0$ we would have to require $u$ to grow fast enough for
large values of its argument, since the propagator would have a
behaviour: $\sim [k u^{-1}(\frac{k^2}{M^2})]^{-1}$.  If the manifold
where the action is defined is such that it allows for large gauge
transformations, like in finite temperature~\cite{Dunne}, this would reduce the
symmetry group, by imposing extra constraints on the gauge
transformation parameter.  Indeed, an $u$ which grows fast with
momenta implies the existence of higher derivatives in the parity
breaking piece of the action. This means that the gauge variation of
the action will involve higher derivatives of the gauge field parameter,
which produce new constraints under the requirement of large gauge
invariance (for example, by integrating by parts the gauge variation).

\section*{Acknowledgements}

This work is partially supported by CONICET (Argentina),
by ANPCyT through grant  No.\ $03-03924$ (AL), and by Fundaci{\'o}n Antorchas (Argentina).

We acknowledge Prof. W.~Bietenholz for kindly pointing out some 
typographical errors and missing  references.

\newpage
%%%%%%%%%%%%%%%%%%%%%%%%%%%%%%%%%%%%%%%%%%%%%%%%%%%%%%%%%%%%%%%%%%%%%%%%%
%%%%%%%%%%%%%%%%%%%%%%%%%%% References %%%%%%%%%%%%%%%%%%%%%%%%%%%%%%%%%%
%%%%%%%%%%%%%%%%%%%%%%%%%%%%%%%%%%%%%%%%%%%%%%%%%%%%%%%%%%%%%%%%%%%%%%%%%


\begin{thebibliography}{bib}
\bibitem{berr}F.~Berruto,~M.~C.~Diamantini and P.~Sodano, {\em `On
    Pure Lattice Chern-Simons Gauge Theories'\/}; hep-th/0004203.
\bibitem{lusch}M.~L\"{u}scher, Phys. Lett. B428, 342 (1998).
\bibitem{anom}For a general reference on anomalies see, for example:
S.~B.~Treiman, R.~Jackiw, B.~Zumino and E.~Witten, {\em Current Algebra
and anomalies}, World Scientific, Singapore 1985, and references
therein.
\bibitem{mcs}S.~Deser, R.~Jackiw and S.~Templeton, Ann.~of Phys.~(N.Y.)
{\bf 140} 372 (1982).
\bibitem{gwr}P.~Ginsparg and K.~Wilson, Phys.~Rev.~{\bf  D25}, 2649
(1982).
\bibitem{biet}W.~Bietenholz and J.~Nishimura,
{\em `Ginsparg-Wilson fermions in odd dimensions'}, hep-lat/0012020.
\bibitem{cicc}M.~L.~Ciccolini, C.~D.~Fosco and F.~A.~Schaposnik,
Phys.~Lett.~{\bf B492}, 214 (2000).
\bibitem{itz}W.~Pauli and F.~Villars, Rev.~Mod.~Phys. {\bf 21}, 434 (1949).
For a modern description of the method, see for example:\\
C.~Itzykson and J.~B.~Zuber, {\em Quantum Field Theory}, Mc Graw-Hill, 1986.
\bibitem{kogan}I.~I.~Kogan, Phys.~Lett.~B {\bf 262}, 1, 83 (1991).
\bibitem{Dunne}For a review, see for example:
 G.~V.~Dunne, Lectures at the 1998 Les Houches Summer School:
  {\em Topological Aspects of Low Dimensional Systems}, e-print arXiv:
  hep-th/9902115.
%%%%%%%%%%%%%%%%%%%%%%%%%%%%%%%%%%%%%%%%%%%%%%%%%%%%%%%%%%%%%%%%%%%%%%%%%%
%%%%%%%%%%%%%%%%%%%%%%%%%%%%%%%%%%%%%%%%%%%%%%%%%%%%%%%%%%%%%%%%%%%%%%%%%%
\end{thebibliography}
\end{document}